\documentclass{aastex}          %% The manuscript based on AASTeX v5.x
\usepackage{spr-astr-addons}    %% mimicing ASTR journal style

\begin{document}

\title{Study of Eclipsing Binary and Multiple Systems in OB Associations V: MQ Cen in Crux OB1}

\shorttitle{Early-type Binary MQ Cen in Crux OB1}
\shortauthors{Bak\i\,{s} et al.}

\author{V. Bak{\i}\c{s}}
\affil{Akdeniz University, Faculty of Science, Department of Space Sciences and Technologies, Antalya, Turkey, email: volkanbakis@akdeniz.edu.tr}
\and
\author{Z. Mikul\'{a}\v{s}ek}
\affil{Masaryk University, Faculty of Science, Department of Theoretical Physics and Astrophysics, Brno, Czech Republic}
\and
\author{J. Jan\'{i}k}
\affil{Masaryk University, Faculty of Science, Department of Theoretical Physics and Astrophysics, Brno, Czech Republic}
%\email{\emaila}
\and
\author{M. Zejda}
\affil{Masaryk University, Faculty of Science, Department of Theoretical Physics and Astrophysics, Brno, Czech Republic}
%\email{\emaila}
\and
\author{E. Tun\c{c}}
\affil{Akdeniz University, Faculty of Science, Department of Space Sciences and Technologies, Antalya, Turkey}
%\email{\emaila}
\and
\author{C. Nitschelm}
\affil{Centro de Astronom{\'i}a (CITEVA), Universidad de Antofagasta, Avenida Angamos 601, Antofagasta, Chile}
%\email{\emaila}
\and
\author{S. Bilir}
\affil{University of Istanbul, Faculty of Science, Department of Astronomy and Space Sciences, Campus, Istanbul, Turkey}
%\email{\emaila}
\and
\author{J. Li\v{s}ka}
\affil{CEITEC, Brno, Czech Republic}
%\email{\emaila}
\and
\author{H. Bak{\i}\c{s}}
\affil{Akdeniz University, Faculty of Science, Department of Space Sciences and Technologies, Antalya, Turkey}

\begin{abstract}
The early-type massive binary MQ~Cen (P$_{orb}$=3.7 d) has been investigated by means of high-resolution ($R\sim48\,000$) spectral analysis and multi-band (Johnson \textit{BVRI} and Str\"{o}mgren $vby$) light curve modeling. The physical parameters of the components have been found to be $M_1= 4.26\pm0.10$~M$_{\odot}$, $R_1= 3.72\pm0.05~$R$_{\odot}$, $T_{\rm eff1}=16\,600\pm520$~K, and $M_2= 5.14\pm0.09~$M$_{\odot}$, $R_2= 7.32\pm0.03~$R$_{\odot}$, $T_{\rm eff2}=15\,000\pm500$~K for the primary and secondary, respectively. The orbital inclination is $i=87.0\pm0.2$ deg. The distance to MQ~Cen has been derived to be $d=2\,460\pm310$ pc which locates it in the Crux OB1 association. However, the age of MQ~Cen ($\sim70$ Myr) is higher than the one reported for the Crux OB1 association ($\sim$6 Myr). The derived masses are implying a spectral type of B5 for the primary and B4 for the secondary component. Nevertheless, the secondary component, which is more massive, appears to be cooler than the primary component: It has completed its lifetime on the main-sequence and it is now positioned at the turn-off point of the giant branch, meanwhile the less massive primary component is still staying on the main-sequence.
\end{abstract}

%% Keywords

%%  Please use labels (\label, \ref) for section, figure, table,
%%  equation  reference. Use \cite for bibliography references.
%
%\section{}%\label{s:?}
%\subsection{}%\label{ss:?}
%\subsubsection{}%\label{sss:?}

\section{Introduction}

Binaries in stellar associations are unique tools, as far as the determination of their astrophysical parameters is concerned. In an effort for identifying eclipsing binary members of stellar associations, several known binary systems in the direction of selected associations have been observed and analyzed with modern analysis techniques \citep[see][]{bakis11, bakis14}. Indeed, observational studies have shown that the multiplicity rate in OB associations is high \citep{bakis15}, as predicted by theoretical studies \citep[see for instance][]{kouwenhoven07}.

The binary star MQ~Cen and the stellar association Crux~OB1 subjected to the present study are located within the borders of the constellation Centaurus. The region partitioned by \citet{humphreys} has been presented in Fig.~1 with the open clusters in the same direction, whose distances and ages range between 780-3400\,pc and 4.3\,Myr - 1\,Gyr \citep{kharchenko05}, respectively. The mean distance of Crux~OB1 is estimated by \citet{kaltcheva} as 2700~pc. The historical studies in the region are very well summarized by \citet{tovmassian} who remarked several B associations up to the distances of 850 pc and several O associations between 1200~pc and 4000~pc for the same direction.

\begin{figure*}
	\begin{center}
		\resizebox{100mm}{!}{\includegraphics{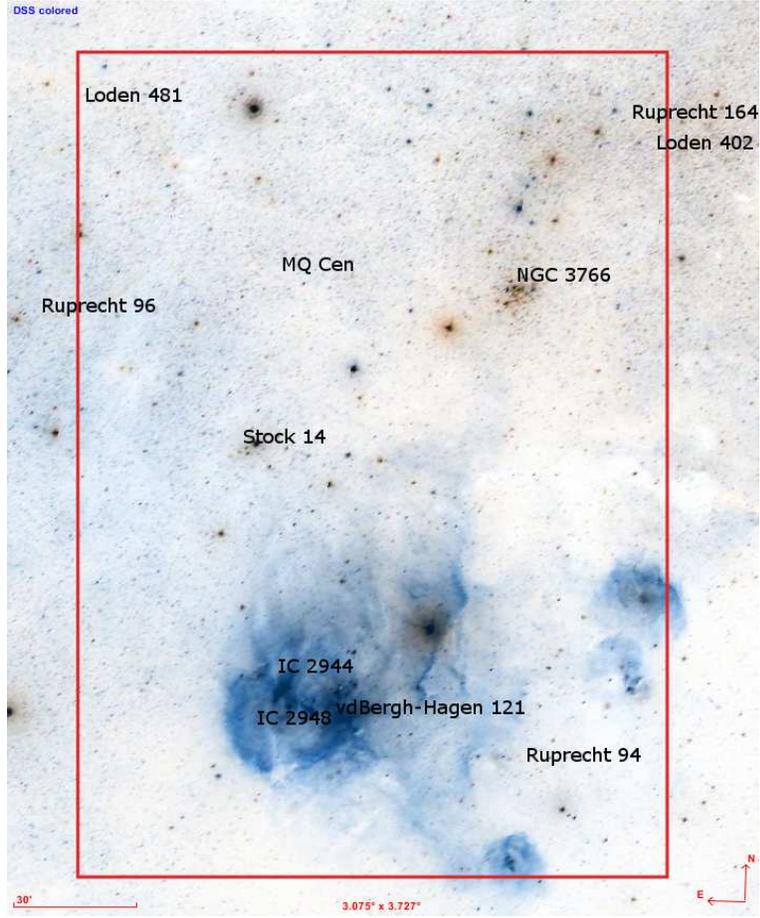}}
		\vspace{.2cm}
		\caption{Crux~OB1 region partitioned by \citet{humphreys}
			between Galactic longitudes \textit{l}=293$^{\circ}$.5 to 295$^{\circ}$.9 and between Galactic latitudes \textit{b}=--2$^{\circ}$.4 to +0$^{\circ}$.1.}\label{fig1}
	\end{center}
\end{figure*}

Having a secondary component at the turn-off point of the main-sequence, MQ~Cen is an interesting eclipsing binary located in the direction of the Crux~OB1 association. The photometric variations of MQ Cen (HD 309074, $V$=10.2\,mag, $\alpha_{2000}=11^{\rm h}\,44^{\rm m}\,16^{\rm s}$, $\delta_{2000}=-61^{\circ}\,42'\,59''$) were discovered by \citet{guthnick} on photographic plates with an amplitude of 0.4\,mag. Since its discovery, there have been a few studies where the times of minimum light of the primary eclipses have been presented \citep[e.g.,][]{dvorak,ogloza}. However, it seems almost impossible to analyse the O--C diagram due to insufficient data. \citet{wolf} published $uvby-\beta$ photometric survey of southern eclipsing binaries including MQ~Cen and its color index. Up to now, to our knowledge, there has not been other study in the literature on the determination of the physical properties of the system.

\section{Observational Data}

\subsection{Photometry}\label{Photometry}

In this study, the new observations have been combined with collected photometric data from several sources (see Table \ref{table1}). Our observations were performed during three seasons using the CELESTRON CGE 1400 XLT 0.35\,m telescope at the South African Astronomical Observatory (SAAO), Sutherland. A Moravian Instrument G2-4000 model CCD camera equipped with \textit{vby} Str\"{o}mgren filters was attached to telescope. In one season, a Cassegrain 762/11430 with a Moravian Instruments CCD camera G2-402 and \textit{BVRI} Johnson filters were used. These observations were conducted mainly by JJ and JL. In total, 13\,546 individual measurements were taken from 2000 to 2017 (see Table \ref{table1} for details).

The data reduction and extraction of the magnitude for each photometric filter were performed by using the {\sc C-Muniwin}\footnote{http://c-munipack.sourceforge.net/} software. The software uses standard bias frames, dark subtraction and flat field division for the preparation of the CCD frames, and then performs synthetic aperture photometry to extract the magnitude.

Additional photometric data were obtained from previous surveys in order to be combined with the observations. In this study, photometric data from bright southern stars have been used from the All Sky Automated Survey (ASAS; \citealt{pojmanski}). In this survey a wide-field 200/2.8 instrument was used and provided photometric measurements in Johnson $V$ and $I$-bands with relatively high accuracy for MQ~Cen. New Supernovae data was found in ASAS using a Nikon AF-S NIKKOR 400mm f/2.8G~ED~VR~AF lens with a ProLine PL230 CCD camera from FLI, with back-illuminated CCD sensor \citep{shappee, kochanek}. The last source of photometric data used in this study was collected using the Optical Monitoring Camera on-board the ESA mission, \textit{INTEGRAL} \citep{OMC}. The instrument has an aperture of 50/154 mm and a CCD camera with a EEV 47-20 CCD chip used for photometry in the \textit{V}-band.

\begin{table*}
	\begin{center}
		\fontsize{10pt}{10pt}\selectfont
		\caption{Photometric data used in this study.}\label{table1}
		\begin{tabular}{lccc}
			\hline\hline
			Source & Filter & Time interval & Number of obs. \\
			\hline
			ASAS   & \textit{V}      & 51878.8-55048.5 & 523 \\
			       & \textit{I}      & 51869.9-53011.8 & 211 \\
			ASAS-SN & \textit{V}     & 57420.7-57926.6 & 718 \\
			OMC     & \textit{V}     & 52818.4-57781.2 & 375 \\
			SAAO    & \textit{v}     & 55694.4-56381.2 & 2295 \\
			& \textit{b}     & 55694.4-56381.2 & 2388 \\
			& \textit{y}     & 55694.4-56381.2 & 2366 \\
			& \textit{B}     & 55297.3-55307.7 & 1074 \\
			& \textit{V}     & 55294.4-55307.7 & 1252 \\
			& \textit{R}     & 55294.4-55307.7 & 1153 \\
			& \textit{I}     & 55294.4-55307.7 & 1191 \\
			\hline
		\end{tabular}
	\end{center}
\end{table*}

All the model parameters have been found by using a weighted robust regression method which eliminated relatively frequent outliers \citep{mik03}. The photometric observations and analysis have given a linear ephemeris of the MQ Cen binary system:

\begin{equation}\label{ephem}
\mathrm{Min I (HJD)}=2\,455\,696.4485(22)+3.686\,963\,1(9) \times E.
\end{equation}

The stability of the orbital period was tested and it was ascertained that $\dot{P}=2(3)\times10^{-9}$. Due to the small value obtained, when compared to the measured uncertainty (see Eq.~1), it could be pertinent to neglect it. The same argument can also be considered valid with the orbital eccentricity whereby because the secondary minimum light occurs at the orbital phase 0.50003(5), the orbital eccentricity can be considered zero.

\subsection{Spectroscopy}

The spectroscopic observations of MQ~Cen were made with the FEROS spectrograph on the 2.2\,m MPG/ESO telescope in La Silla, Chile. During five observing nights in March 2011, a total of 25 high-resolution $(R (\lambda / \Delta\lambda) \sim48\,000)$ spectra were collected. FEROS is fed by two fibers and each fiber is illuminated via 2.0~arcsec aperture, which means that the third light source detected in the photometry is not included in the composite spectra. The exposure times were set to 1200 s to obtain an average of S/N ratio of $\sim$120 at 6000 {\AA}. The {\sc IRAF} package was used for continuum normalization of the data reduction pipeline. For each telescope pointing, two consecutive exposures were taken to allow an estimation of the spectrum S/N ratio. This is accomplished by investigating the product of the division between the two consecutive spectra. Through this methodology, any spectral feature which originates from either the stellar atmospheres or from the interstellar medium, will cancel each other, leaving only the observational bias. In this way, the S/N ratio has been calculated for each spectrum at six different wavelengths to check the relative uncertainties in RV measurements.

Spectral lines of both components are clearly visible in the spectra, which allow the double-lined radial velocity curves of the system to be obtained. To disentangle spectral lines of both components, and to determine their radial velocities we used the code {\sc korel} \citep{hadrava} in {\sc vo-korel} implementation\footnote{see \url{https://stelweb.asu.cas.cz/vo-korel/}} \citep{skoda}. We have selected three spectral regions covering four strong spectral lines He 4471 {\AA} + Mg 4481 {\AA}, He 5875 {\AA}, and He 6678 {\AA}. The radial velocities and their uncertainties are given in Table \ref{table2}. Spectroscopic orbital solution obtained from {\sc korel} was combined with the fixed orbital period \textit{P} given in Eq. \ref{ephem} and by setting the eccentricity to zero (\textit{e} = 0) resulted in following values: $K_{1} = 158.9\pm0.8$ km s$^{-1}$, $K_{2} = 131.6\pm0.7$ km s$^{-1}$, $q = K_1/K_2 = 1.207\pm0.009$ and $V_\gamma = -1.2\pm4.8$ km s$^{-1}$.

It is known that the broad, overlapping spectral lines may cause incorrect orbital velocity determinations up to 2-3 per cent \citep[e.g.][]{popper}. This is the result of line core shifts when two broad lines overlap or blend, so that traditional RV measuring techniques like cross-correlation (e.g., fitting a Gaussian to line cores), gives an incorrect velocity. However, for modern techniques like spectral disentangling, the spectral lines of the components move along the wavelength axis to match the composite spectrum. In this case, it removes the effects from the overlapping of spectral lines in RV measurements, which makes it a very efficient tool for orbital parameter determination of double-lined spectroscopic binaries.

\begin{figure*}
	\begin{center}
		\begin{tabular}{cc}
			\resizebox{80mm}{!}{\includegraphics{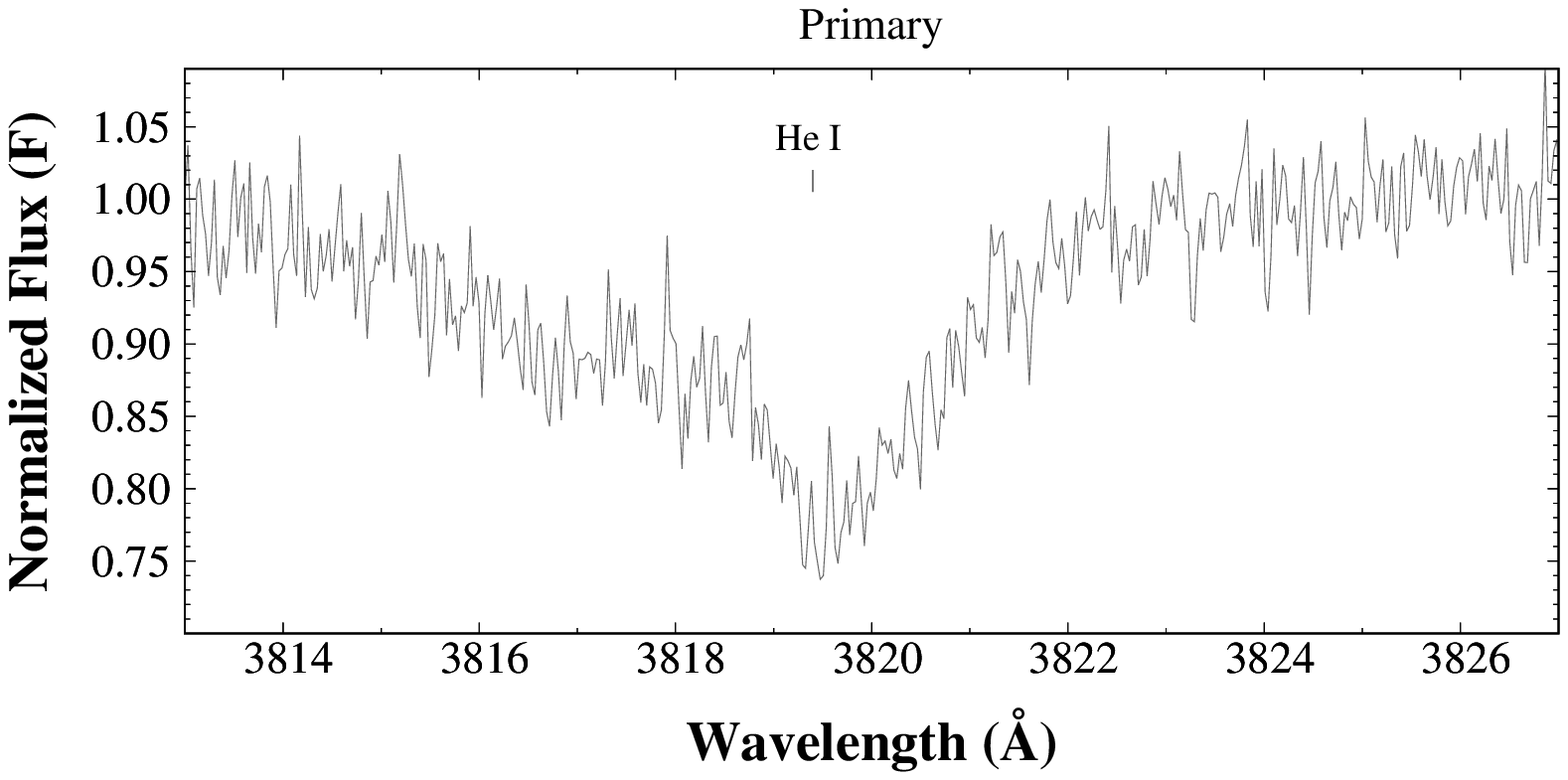}} & \resizebox{80mm}{!}{\includegraphics{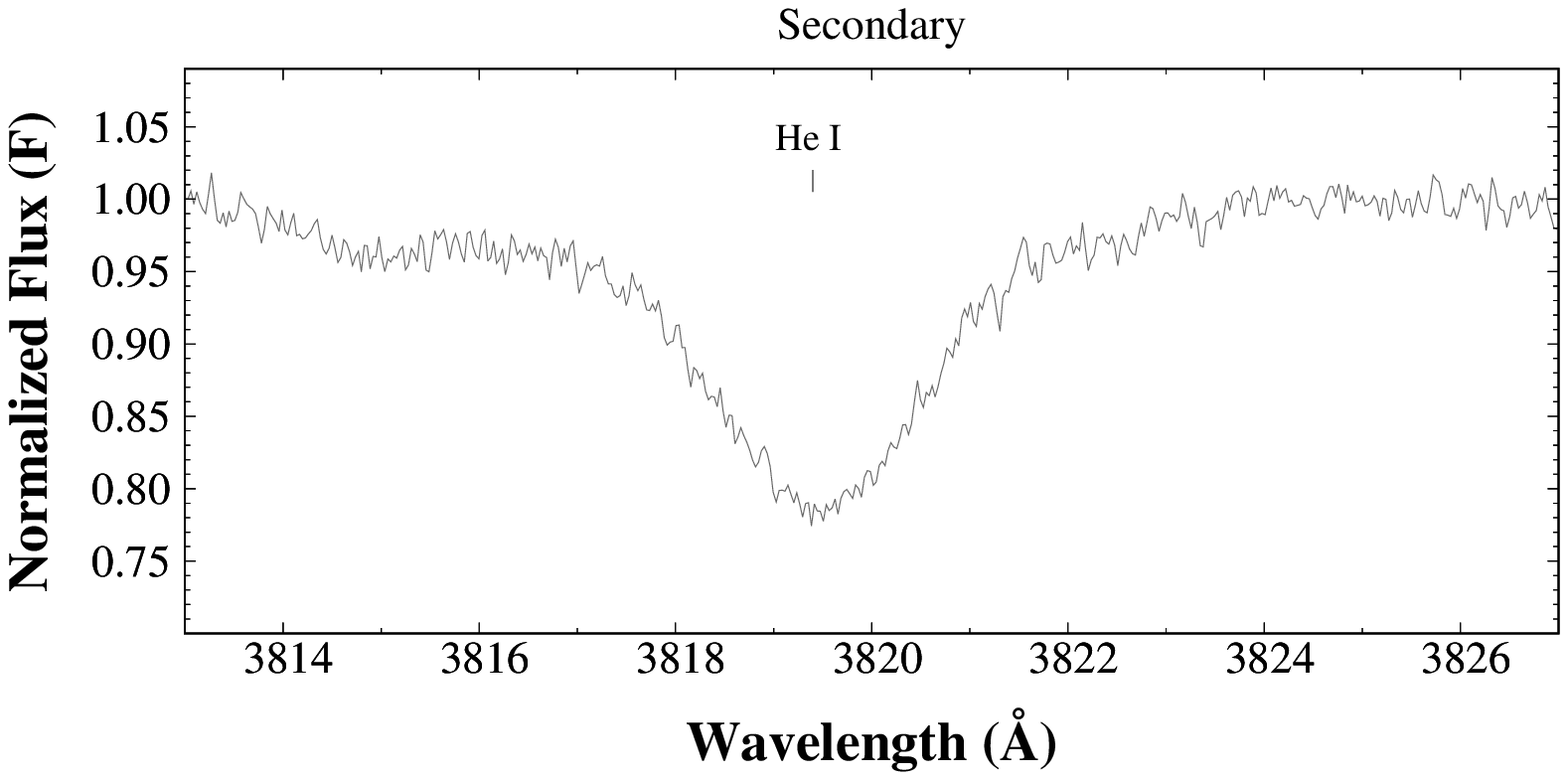}}\\
			\resizebox{80mm}{!}{\includegraphics{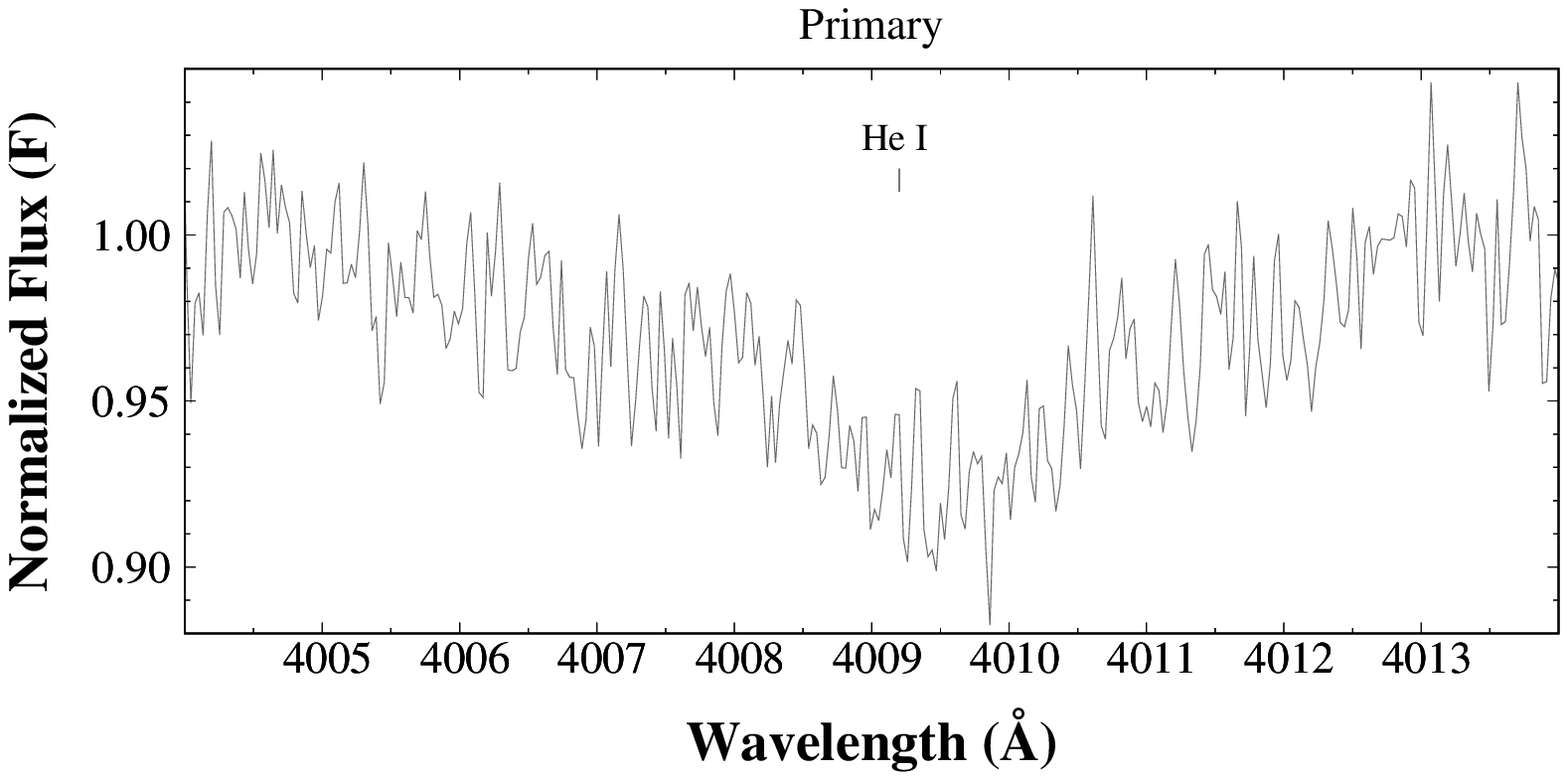}} &
			\resizebox{80mm}{!}{\includegraphics{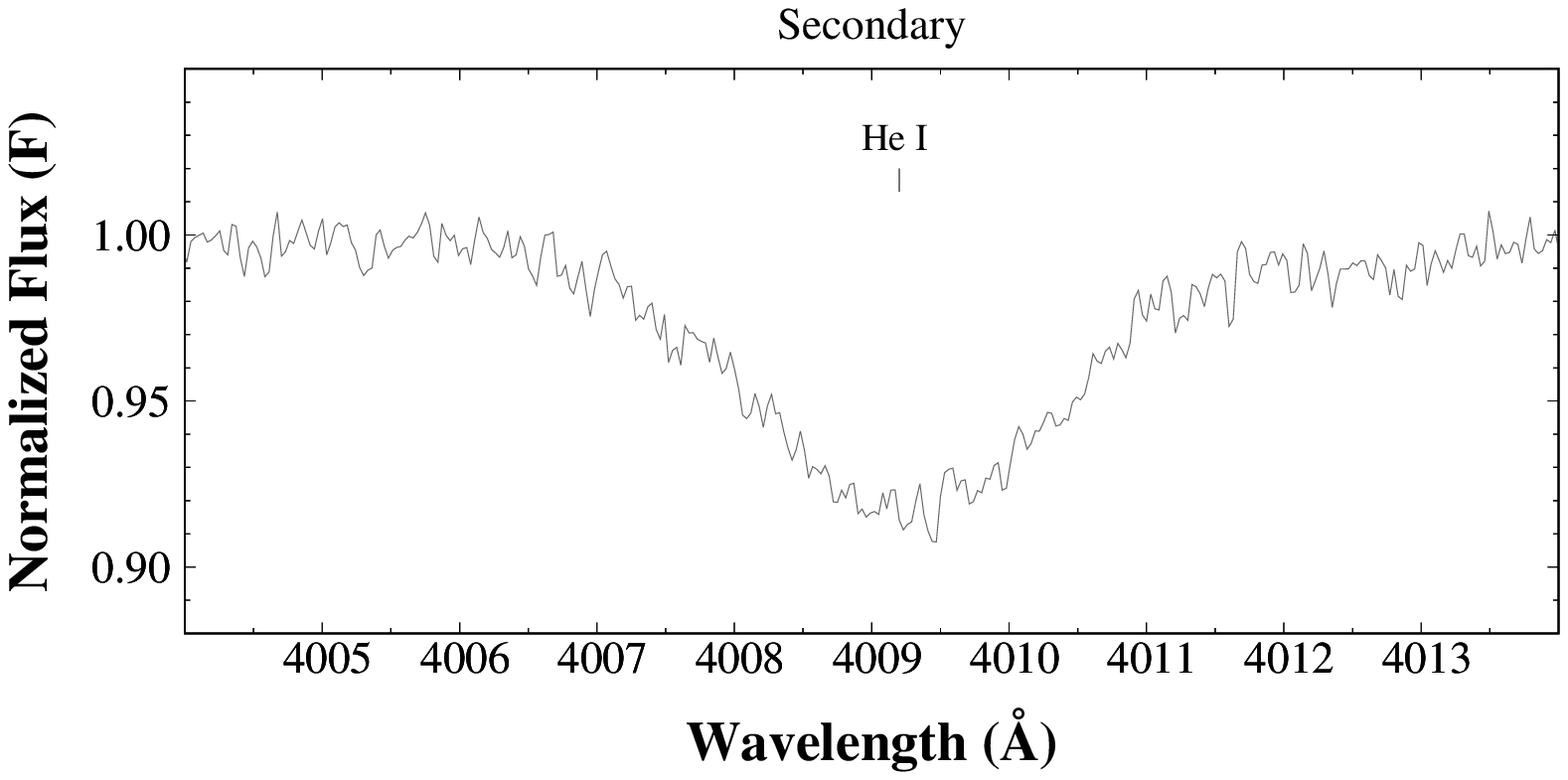}}\\
			\resizebox{80mm}{!}{\includegraphics{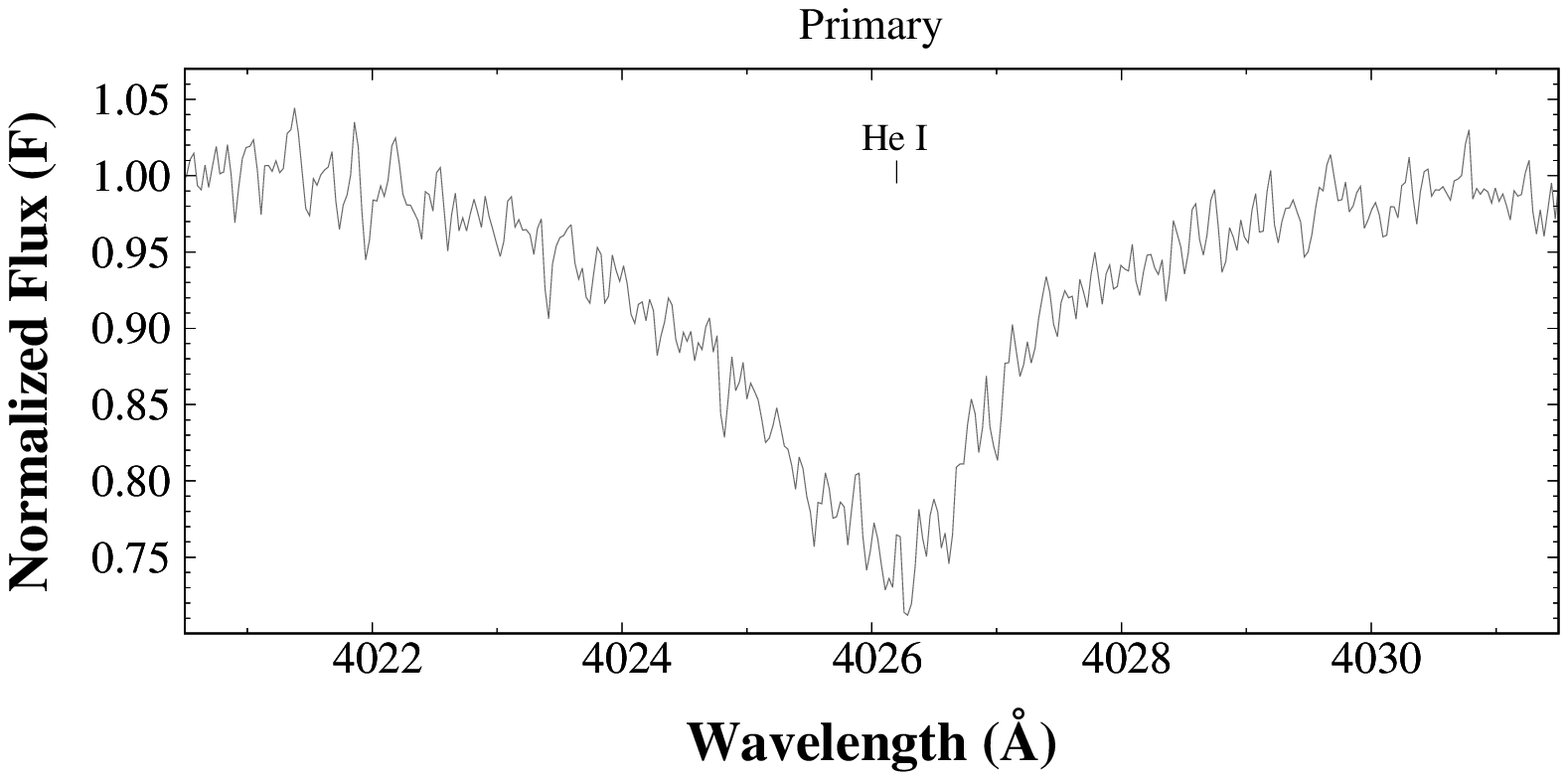}} &
			\resizebox{80mm}{!}{\includegraphics{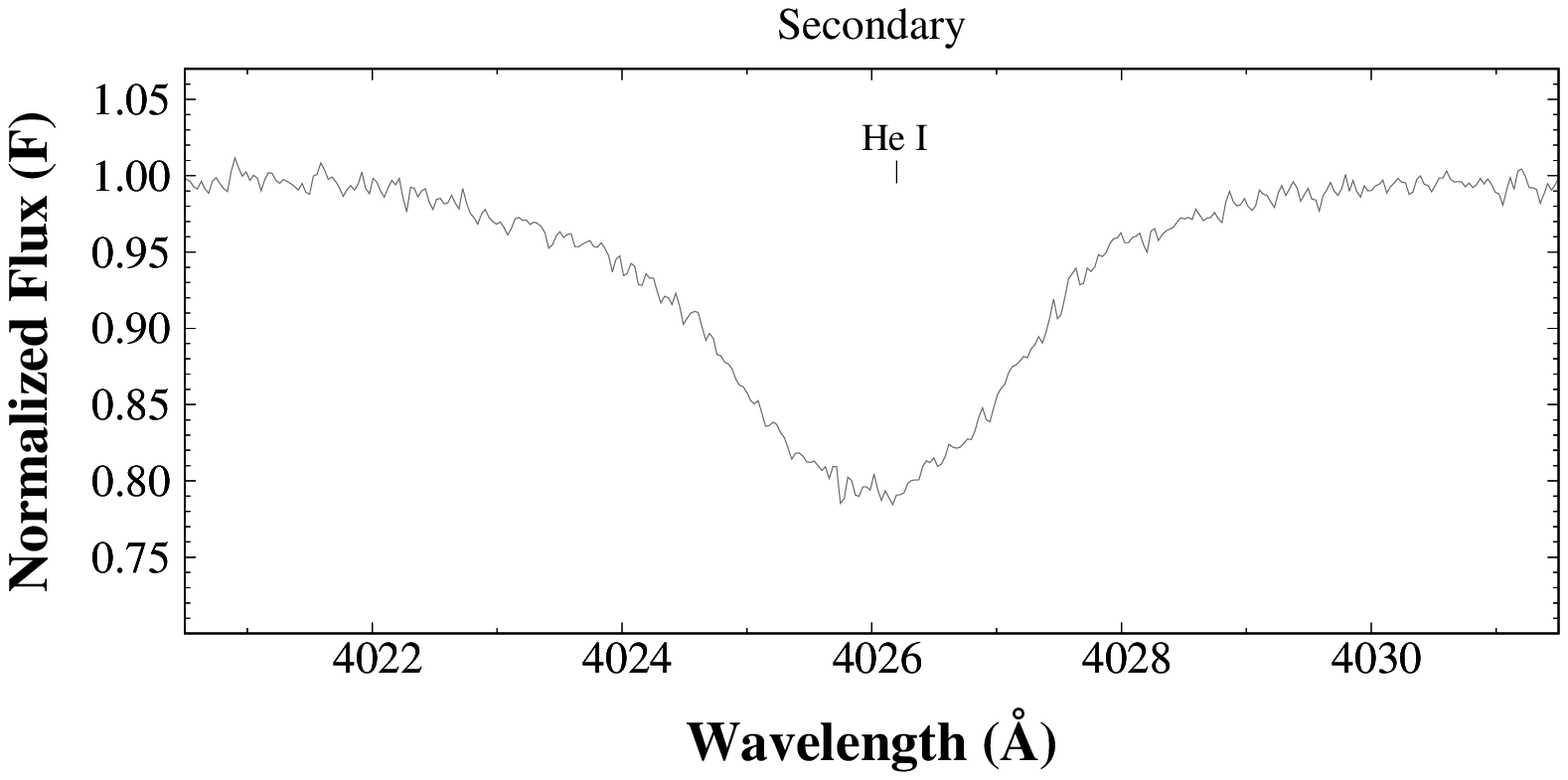}}\\
			\resizebox{80mm}{!}{\includegraphics{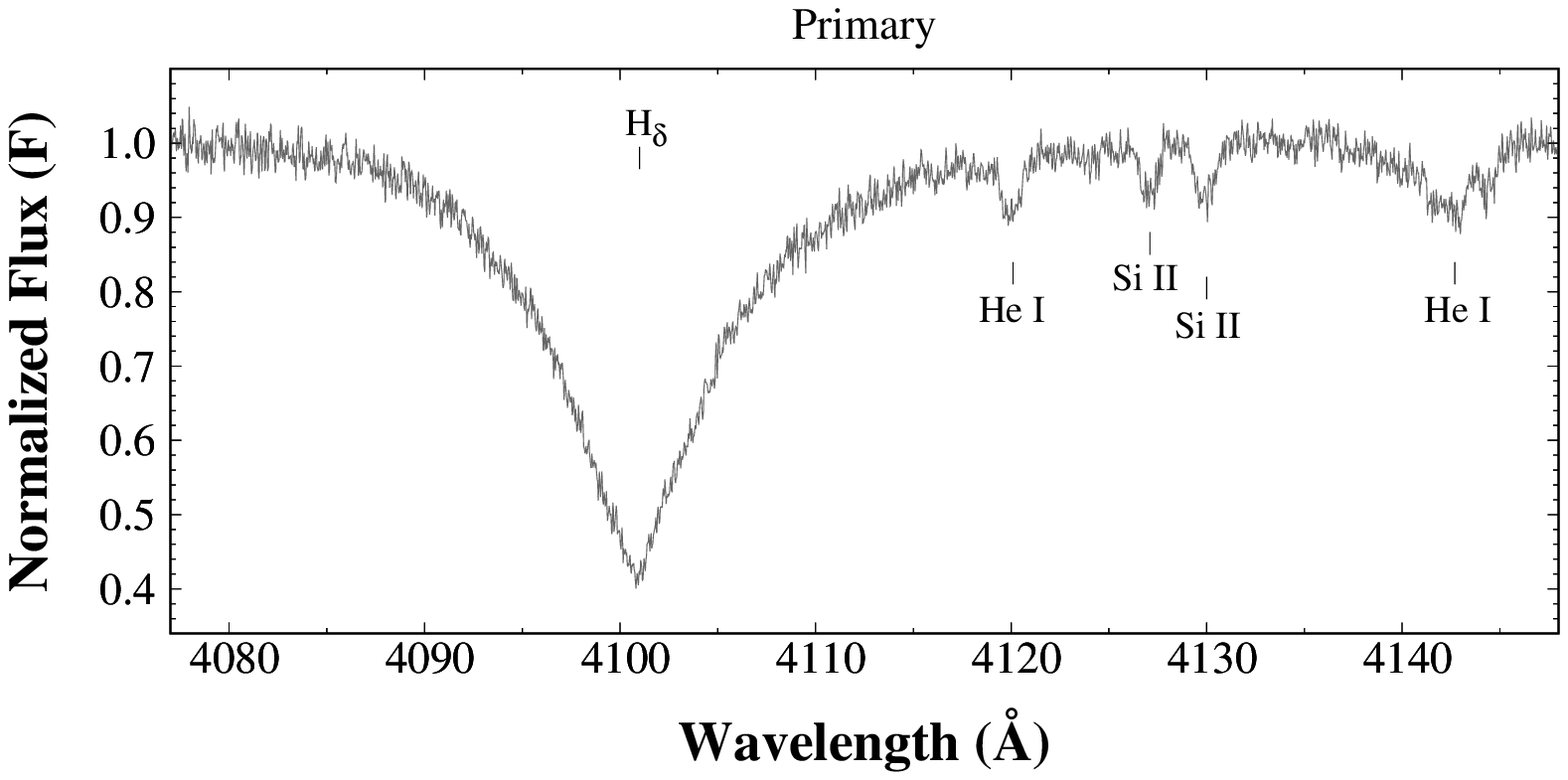}} &
			\resizebox{80mm}{!}{\includegraphics{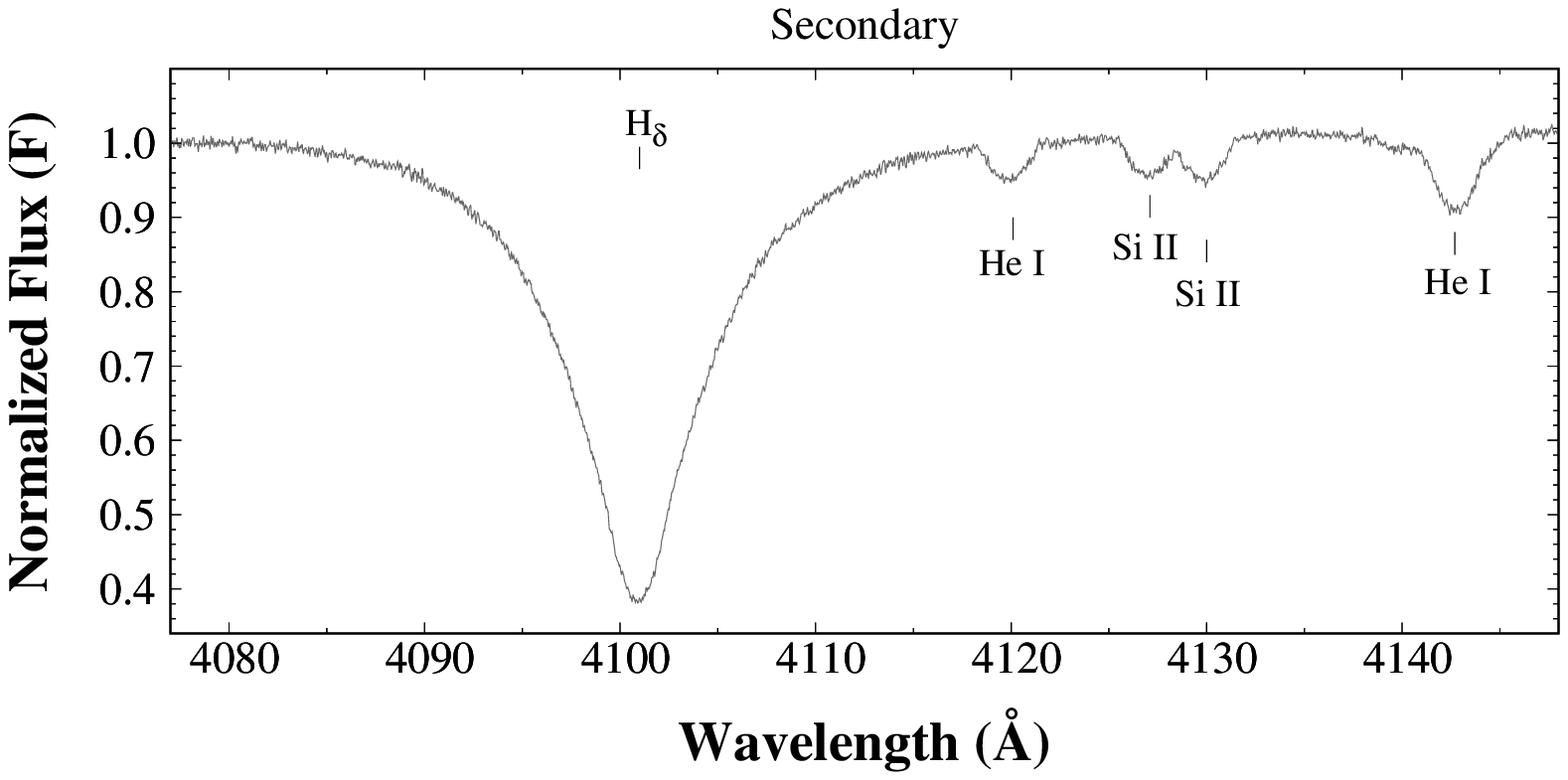}}\\
			\resizebox{80mm}{!}{\includegraphics{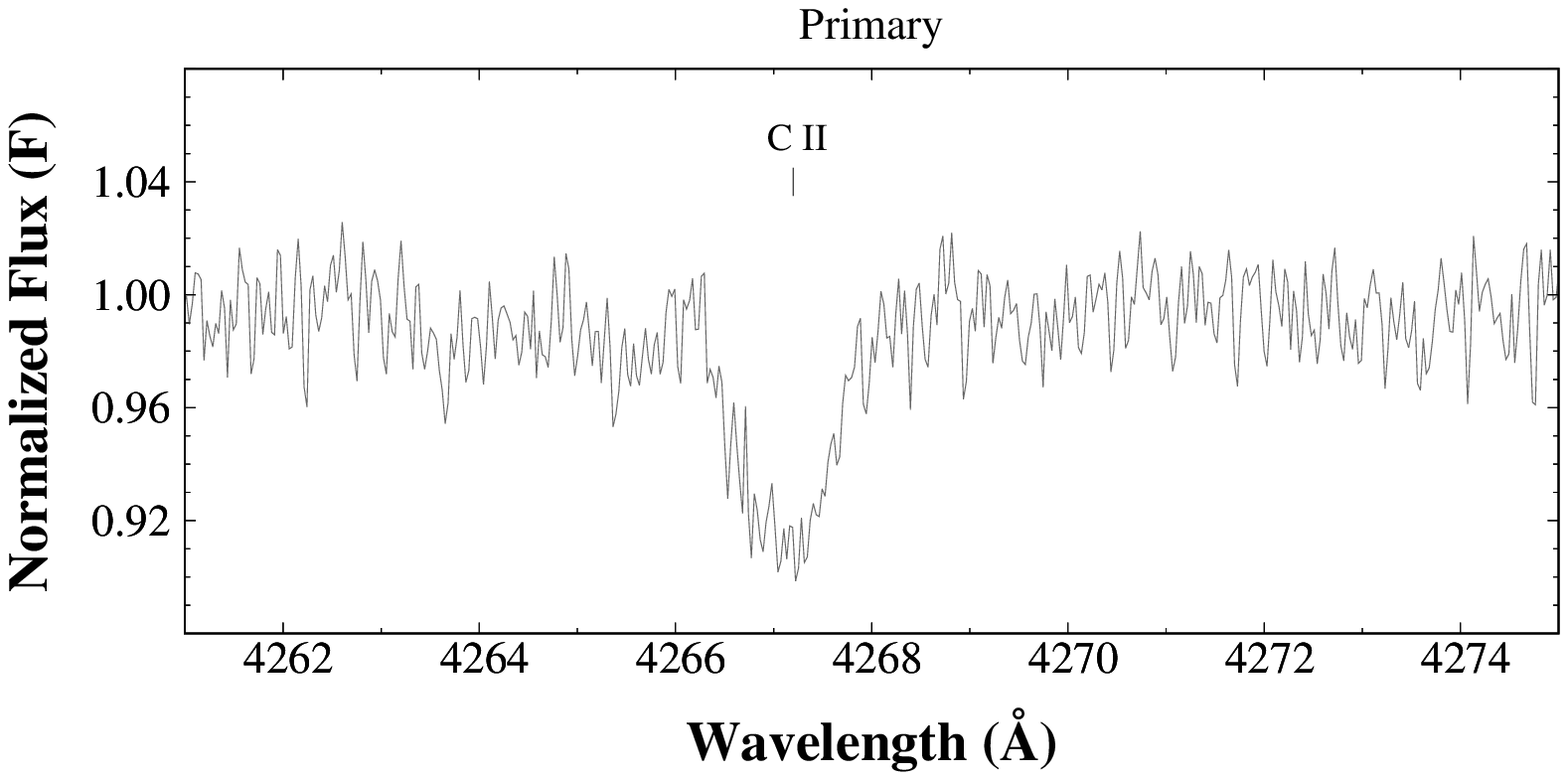}} &
			\resizebox{80mm}{!}{\includegraphics{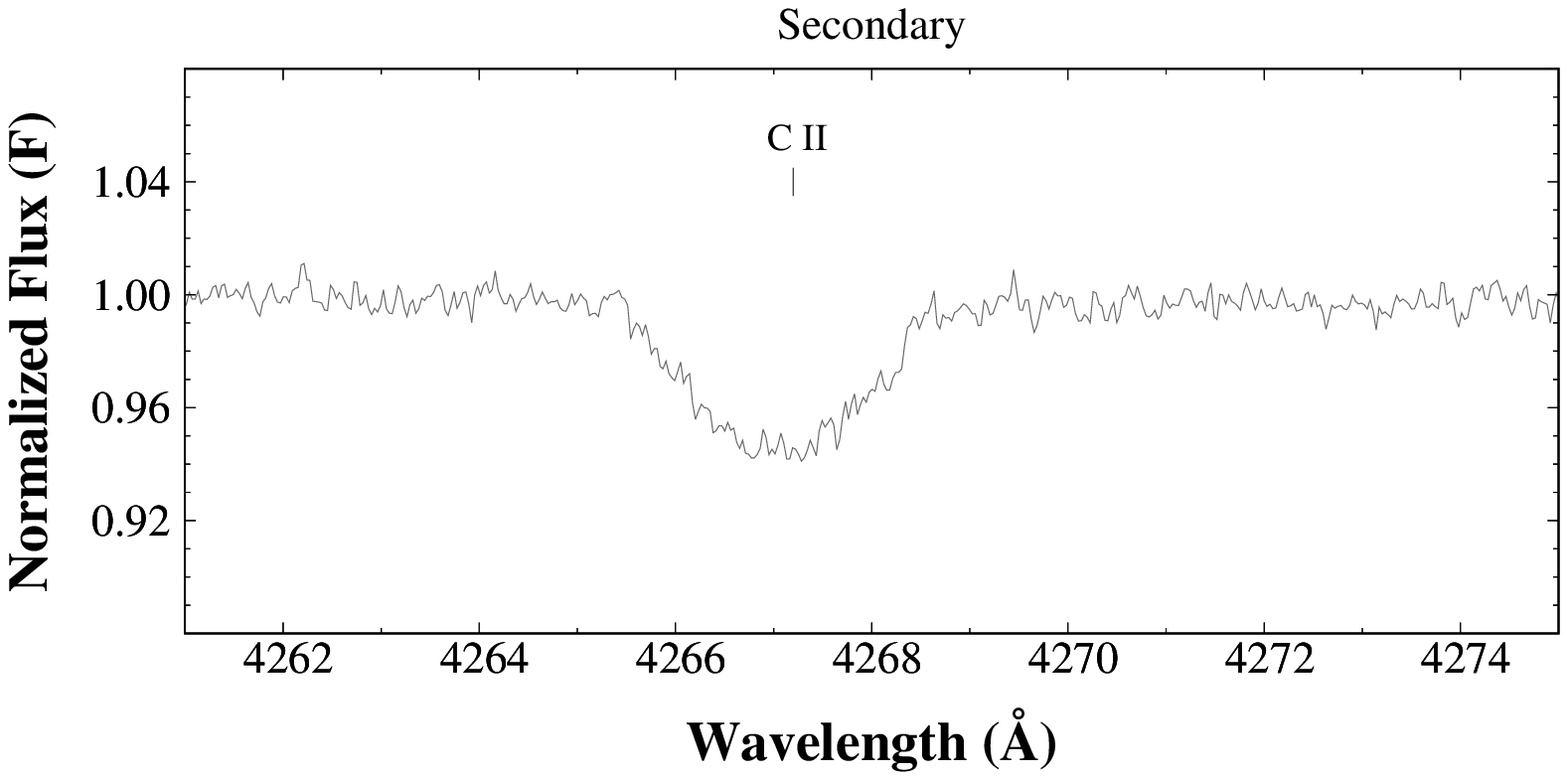}}\\
			%			\vspace{0.5cm}
		\end{tabular}
		\caption{Disentangled spectral lines of MQ Cen components.}
		\label{fig2}
	\end{center}
\end{figure*}

\begin{table*}
	\begin{center}
		\fontsize{8pt}{10pt}\selectfont
		\caption{Journal of spectroscopic observations.}
		\label{table2}
		\begingroup
		\setlength{\tabcolsep}{4pt} % Default value: 6pt
		\renewcommand{\arraystretch}{1.5} % Default value: 1
		\begin{tabular}{lllcccccccccccc}
			\hline\hline
			ID &Observation &Phase & \multicolumn{4}{c}{HeI4471 {\AA} + MgII4481 {\AA}} & \multicolumn{4}{c}{HeI5875 {\AA}} & \multicolumn{4}{c}{HeI6678 {\AA}} \\
			&(HJD - & ($\phi$) & RV$_{1}$ & err & RV$_{2}$ & err & RV$_{1}$
			& err & RV$_{2}$ & err & RV$_{1}$ & err & RV$_{2}$ & err \\
			&2\,400\,000)&  & \multicolumn{2}{c}{km s$^{-1}$}
			& \multicolumn{2}{c}{km s$^{-1}$} & \multicolumn{2}{c}{km s$^{-1}$} &
			\multicolumn{2}{c}{km s$^{-1}$} & \multicolumn{2}{c}{km s$^{-1}$} &
			\multicolumn{2}{c}{km s$^{-1}$} \\
			\hline
			\,\,1 & 55647.6116&  0.753 & 158.0  & -0.7 & -132.4 & -0.8 & 156.3  & -2.4 & -133.7 & -2.2 & 156.6  & -2.2 & -138.8 & -7.3  \\
			\,\,2 & 55647.6261&  0.757 & 158.6  & 0.1  & -131.8 & -0.5 & 156.3  & -2.3 & -133.7 & -2.4 & 156.5  & -2.0 & -132.3 & -0.9  \\
			\,\,3 & 55648.4967&  0.993 & 14.5   & 11.0 & -2.3   & 0.6  & 5.7    & 2.3  & -1.8   & 1.0  & 3.2    & -0.3 & -1.9   & 0.9   \\
			\,\,4 & 55648.5112&  0.997 & -1.6   & -1.2 & 0.0 & -0.4 & 2.4    & 2.9  & 1.3    & 0.9  & 2.0    & 2.5  & -0.6   & -1.0 \\
			\,\,5 & 55648.5257&  0.001 & 6.9    & 11.3 & 4.8 & 1.2  & -4.6   & -0.2 & 3.7    & 0.1  & -4.4   & 0.0  & 3.1    & -0.5 \\
			\,\,6 & 55648.5402&  0.005 & -8.8   & -0.5 & 6.9 & 0.0  & -7.4   & 0.9  & 6.4    & -0.5 & -8.7   & -0.4 & 7.0    & 0.1   \\
			\,\,7 & 55648.5564&  0.010 & -7.1   & 5.5  & 10.1 & -0.4 & -13.7  & -1.1 & 10.2   & -0.3 & -13.1  & -0.4 & 10.8   & 0.3   \\
			\,\,8 & 55648.5709&  0.014 & -15.9  & 0.7  & 13.8 & 0.0  & -6.3   & 10.3 & 15.6   & 1.8  & -18.3  & -1.7 & 13.8   & 0.1   \\
			\,\,9 & 55648.5854&  0.017 & -16.1  & 4.4  & 15.9 & -1.1 & -19.0  & 10.5 & 16.6   & -0.4 & -21.2  & -0.7 & 17.8   & 0.8   \\
			10& 55648.5999&  0.021 & -16.7  & 7.7  & 21.3 & 1.1  & -24.0  & 0.4  & 21.2   & 1.0  & -26.2  & -1.8 & 21.2   & 1.0   \\
			11& 55648.6146&  0.025 & -22.3  & 6.1  & 22.2 & -1.2 & -26.7  & 1.6  & 25.8   & 2.3  & -28.4  & -0.1 & 23.7   & 0.3   \\
			12& 55648.6291&  0.029 & -35.7  & -3.6 & 26.1 & -0.6 & -31.4  & 0.8  & 26.3   & -0.3 & -33.6  & -1.4 & 28.7   & 2.1   \\
			13& 55649.5273&  0.273 & -156.5 & 0.2  & 129.9  & 0.0  & -156.6 & 0.2  & 130.9  & 1.1  & -157.1 & -0.4 & 129.6  & -0.3  \\
			14& 55649.5418&  0.277 & -157.5 & -1.4 & 129.9  & 0.7  & -156.2 & -0.2 & 129.0  & -0.3 & -156.2 & -0.2 & 130.1  & 0.8   \\
			15& 55649.6172&  0.297 & -150.3 & 0.6  & 126.0  & 0.9  & -151.2 & -0.3 & 126.1  & 1.0  & -151.2 & -0.3 & 126.2  & 1.1   \\
			16& 55649.6317&  0.301 & -149.9 & -0.2 & 125.9  & 1.9  & -150.1 & -0.4 & 126.2  & 2.2  & -156.3 & -6.6 & 126.2  & 2.2   \\
			17& 55650.4902&  0.534 & 42.0   & 5.2  & -30.0  & 0.5  & 37.4   & 0.5  & -31.2  & -0.7 & 36.9   & 0.0  & -31.2  & -0.7 \\
			18& 55650.5047&  0.538 & 54.0   & 13.4 & -32.8  & 0.8  & 40.6   & -0.1 & -33.6  & 0.0  & 41.2   & 0.6  & -34.2  & -0.6 \\
			19& 55650.5505&  0.550 & 53.4   & 0.9  & -45.1  & -1.7 & 52.7   & 0.2  & -46.0  & -2.5 & 53.1   & 0.6  & -43.7  & -0.3 \\
			20& 55650.5650&  0.554 & 57.9   & 1.7  & -46.9  & -0.3 & 55.2   & -1.0 & -47.1  & -0.6 & 56.2   & 0.1  & -48.0  & -1.5 \\
			21& 55650.7014&  0.591 & 89.8   & 0.9  & -73.2  & 0.4  & 89.1   & 0.2  & -73.2  & 0.5  & 88.9   & 0.0  & -73.7  & -0.1 \\
			22& 55650.7159&  0.595 & 90.4   & -1.7 & -77.3  & -1.0 & 91.7   & -0.4 & -76.5  & -0.2 & 91.6   & -0.6 & -78.2  & -1.9  \\
			23& 55651.4902&  0.805 & 148.4  & 0.1  & -122.1 & 0.7  & 148.1  & -0.2 & -123.0 & -0.2 & 148.6  & 0.4  & -122.0 & 0.9  \\
			24& 55651.5048&  0.809 & 146.1  & -0.7 & -121.8 & -0.2 & 147.0  & 0.1  & -122.9 & -1.3 & 146.6  & -0.2 & -121.5 & 0.1   \\
			25& 55651.5941&  0.834 & 135.9  & -0.1 & -112.8 & -0.2 & 135.8  & -0.1 & -112.6 & 0.0  & 136.1  & 0.2  & -117.8 & -5.2 \\
			\hline
		\end{tabular}
		\endgroup
	\end{center}
\end{table*}

The individual spectrum of the components was obtained by using the {\sc korel} code, with the orbital parameters derived in Section\,\ref{Photometry} and combined with the component light factors for the epoch at each observed spectrum derived from the light curve solution (see \S3). Fig. \ref{fig2} shows a selection of component spectra between $\lambda 3800-4300$\,{\AA}.

\section{Analysis of the Photometric Data}

\begin{figure}
	\begin{center}
		\resizebox{50mm}{!}{\includegraphics{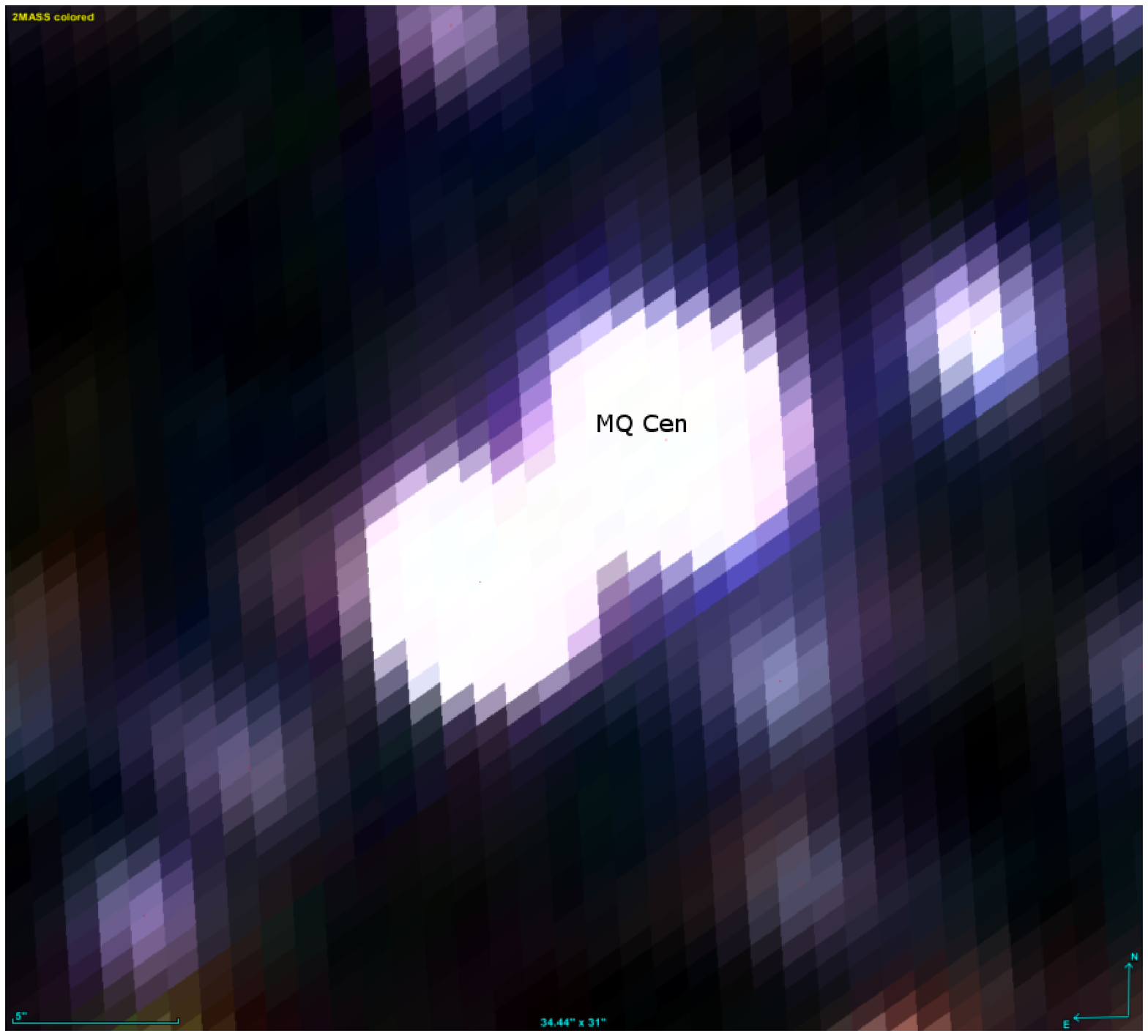}}
		%		\vspace{.5cm}
		\caption{MQ~Cen and the third light source at 12'' South-East. The image dimension is 34'' $\times$ 31''.}\label{fig3}
	\end{center}
\end{figure}

\begin{figure*}
	\begin{center}
		\resizebox{80mm}{!}{\includegraphics{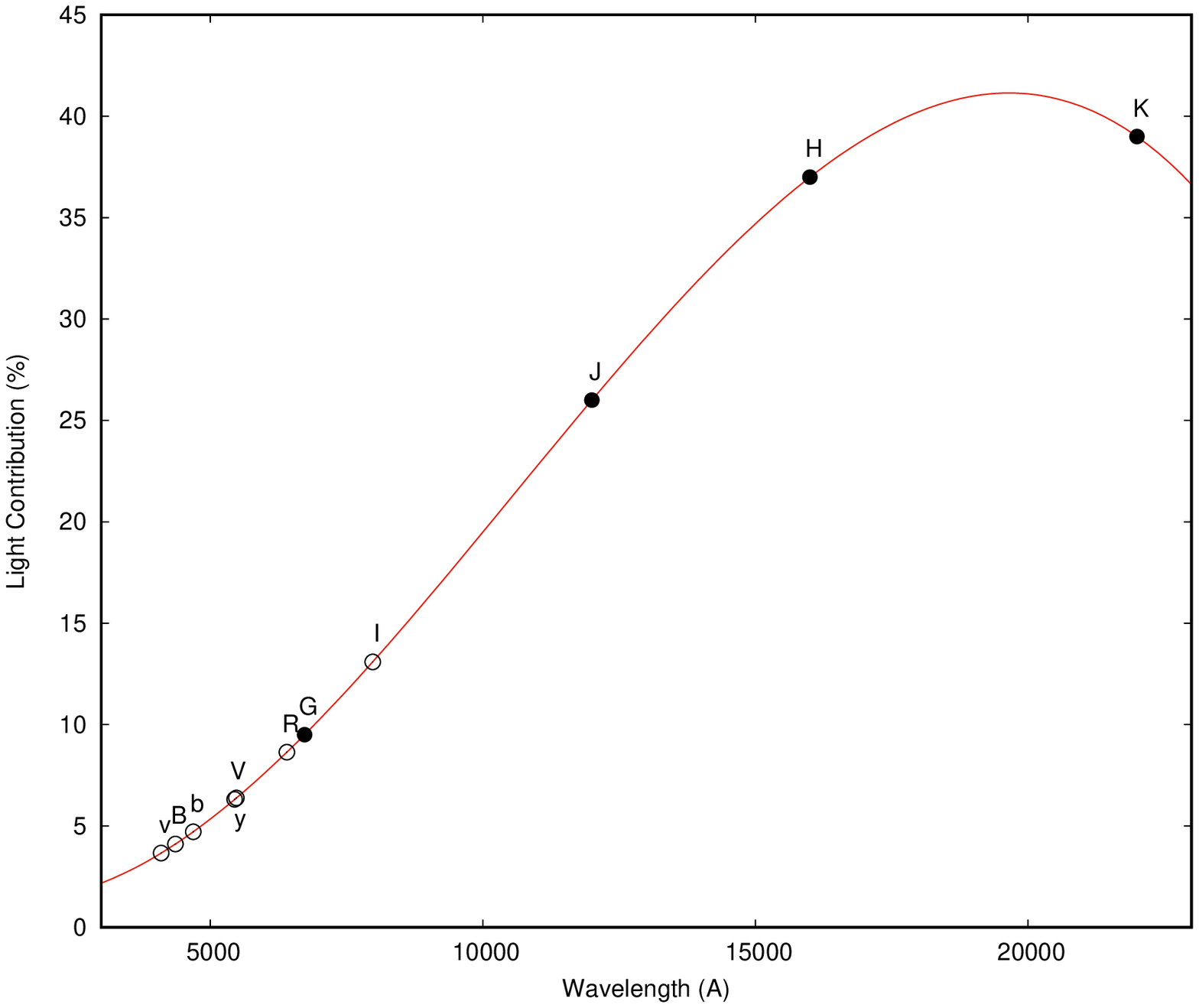}}
		%		\vspace{.2cm}
		\caption{In percentage, light contribution of the visible companion versus wavelength. The observations are represented by filled circles, while, the open circles represent the calculated values obtained from the polynomial fit. For each data point the spectral band label is given.}\label{fig4}
	\end{center}
\end{figure*}

Prior to analysis of the photometric data, it was checked to see if there were any light sources within 30~arcsec of MQ~Cen in the All-Sky Catalogue of Point Sources (2MASS) Collaboration catalogue \citep{cutri}. We found an object, 2MASS J11441637-6143033, at 12~arcsec south-east of MQ~Cen (Fig. \ref{fig3}). The magnitude differences between MQ~Cen and the third light source in $J$, $H$ and $K_s$-bands are 1.13, 0.58 and 0.50 mag, respectively, which correspond to 26\,\%, 37\,\% and 39\,\% light contribution to the total flux. MQ~Cen and its visual companion are observed by the \textit{Gaia} \citep{gaia} satellite. The magnitude difference in \textit{Gaia} \textit{G}-band is 2.448 mag which corresponds to 10\,\% light contribution of the visual companion to the total flux. Using the light contributions from \textit{G, J, H} and \textit{K}-bands, a third order polynomial function was fitted to obtain the light contributions for other bands used in our study (Fig. \ref{fig4}).

The multi-band photometric data described in \S2.1 were modeled using the 2003 version of the Wilson-Devinney light curve and radial velocity analysis code \citep[][]{wilson}. The code requires the temperature of one of the components as a priori to calculate the temperature of the other component. A more reliable and direct way to determine the temperature is through the modeling of the stellar spectrum with theoretical stellar atmosphere models. 

We obtained a spectrum at the orbital phase $\phi\sim$\,0.0, where the less massive but hotter primary component is completely occulted and we only see the more massive but cooler secondary star. 

In our study, a spectrum at the orbital phase $\phi\sim$\,0.0 has been obtained, where the less massive but hotter primary component is completely occulted, and we only see the more massive but cooler secondary star has been observed.

\begin{figure*}
	\begin{center}
		\begin{tabular}{cc}				
			\resizebox{120mm}{!}{\includegraphics{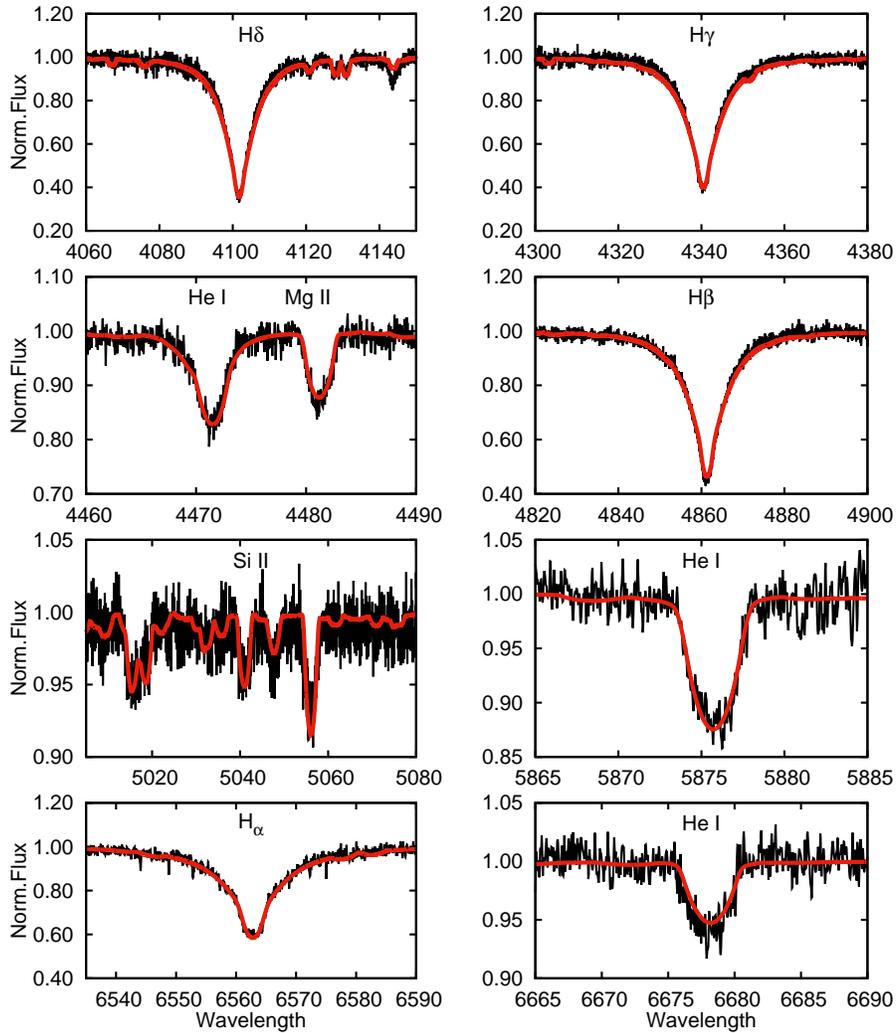}}\\
		\end{tabular}
%		\vspace{1cm}
		\caption{Fitted synthetic spectra of selected spectral lines at the orbital phase $\phi=0.0$ for MQ~Cen.}\label{fig5}
	\end{center}
\end{figure*}

Therefore, this spectrum is suitable for determining the atmospheric parameters such as temperature, surface gravity, projected rotational velocity and micro-turbulance velocity. A grid of \citet{kurucz} model stellar atmospheres and synthetic spectra was created by using the {\sc atlas9} and {\sc synthe} codes, respectively. The best fitting stellar atmosphere models yielded $T_{\rm eff2} = 15\,000\pm500$~K, $\log~g$ = 3.4 (cgs units), $v_{\rm rot}\sin i$ = 100~km s$^{-1}$ for the secondary component. The micro-turbulence velocity was adopted to be $\zeta$ = 2.0~km s$^{-1}$. The uncertainties are the standard deviations of the parameters obtained from each line fitting. Some of the fitted stellar atmosphere models are shown in Fig. \ref{fig5}. The same temperature value has been obtained after using the Str\"{o}mgren colours of the system given by \citet{wolf}.

\begin{figure*}
	\begin{center}
		\begin{tabular}{c}
			\resizebox{100mm}{!}{\includegraphics{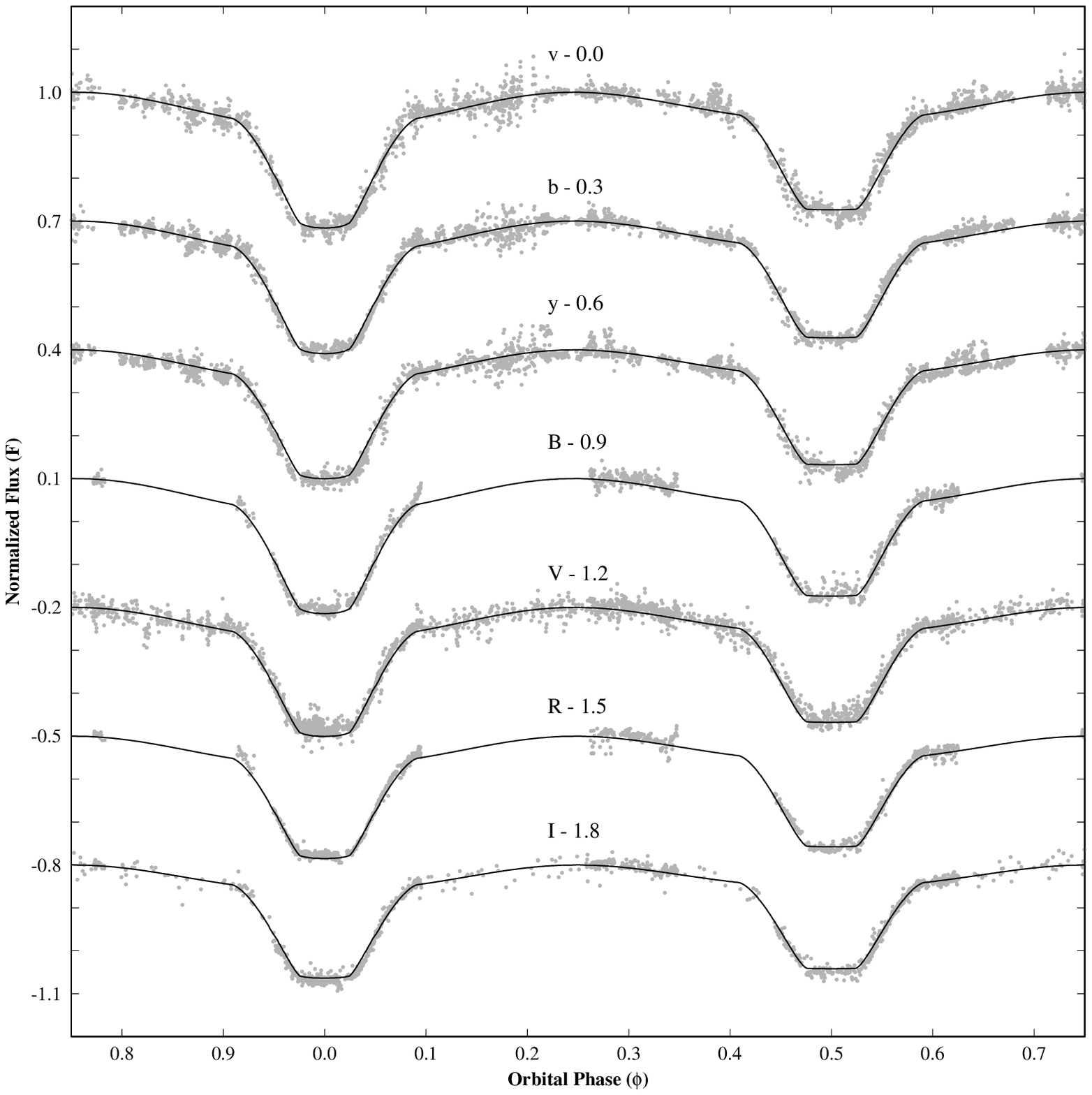}}\\
			\resizebox{100mm}{!}{\includegraphics{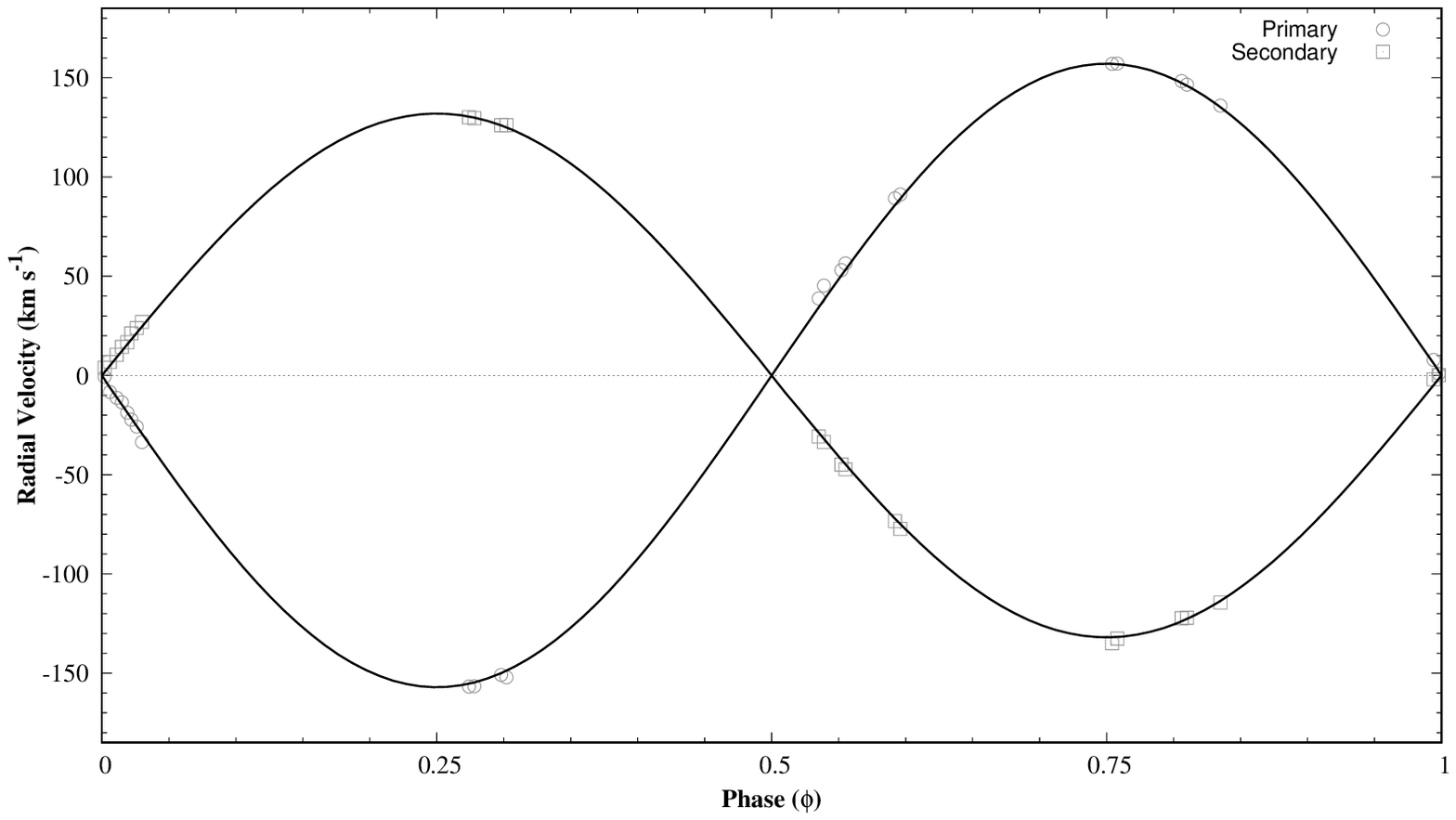}}\\
		\end{tabular}
%		\vspace{1cm}
		\caption{Best fitting models for the light curves (upper panel) and the spectroscopic orbit (lower panel).}\label{fig6}
	\end{center}
\end{figure*}

During the modeling of the light curves, the bolometric albedo and the gravity darkening exponents were set to 1.0 as suggested for early-type stars. The logarithmic Limb Darkening (LD) law was adopted and the LD coefficients were taken from \citet{vanhamme}. The detached configuration was first adopted to see if the components were inside or outside their Roche lobe limits. The light curve parameters conjunction time ($T_0$), orbital inclination ($i$), primary star effective temperature ($T_{\rm eff1}$), surface potentials ($\Omega_{\rm 1,2}$) and light contribution of the primary component ($L_1$) were allowed to vary while the orbital period ($P$), orbital eccentricity ($e$) and secondary star effective temperature ($T_{\rm eff2}=15000\,K$) were fixed. According to the best fit obtained with the light curve model, the orbital inclination indicates that, at total eclipse, the less massive primary component is within its Roche lobe limit while the more massive secondary star is almost filling its Roche lobe limit. The final light curve parameters have been presented in Table \ref{table3} and the model light curve together with the spectroscopic orbit are shown in Fig. \ref{fig6}.

\begin{table*}
	\begin{center}
		\caption{Solution parameters for the light curve and radial velocity curve. Error in brackets quotes uncertainty in the last digit.}
		\label{table3}
		\begin{tabular}{lc}\hline\hline
			Parameter              &  Value \\
			\hline
			$P$ (days)             & \multicolumn{1}{c}{3.6869627(10)}  \\
			$T_0$ (HJD)            & 2455552.6599(3) \\
			$T_{\rm eff1}$(K)      &  \multicolumn{1}{c}{16\,600 (20)} \\
			$T_{\rm eff2}$(K)      &   15\,000   \\
			$q$                    &  \multicolumn{1}{c}{1.207} \\
			$V_\gamma$\,(km s$^{-1})$&  \multicolumn{1}{c}{--1.2} \\
			$L_{1}/L_{1+2}$ ($v$)  &   0.2254(1) \\
			$L_{1}/L_{1+2}$ ($b$)  &   0.2219(1) \\
			$L_{1}/L_{1+2}$ ($y$)  &   0.2155(1) \\
			$L_{1}/L_{1+2}$ ($B$)  &   0.2254(6) \\
			$L_{1}/L_{1+2}$ ($V$)  &   0.2184(5) \\
			$L_{1}/L_{1+2}$ ($R$)  &   0.2084(6) \\
			$L_{1}/L_{1+2}$ ($I$)  &   0.1920(5) \\
			$L_{3}/L_{1+2+3}$ ($v$)  &   0.04(1) \\
			$L_{3}/L_{1+2+3}$ ($b$)  &   0.05(1) \\
			$L_{3}/L_{1+2+3}$ ($y$)  &   0.06(1) \\
			$L_{3}/L_{1+2+3}$ ($B$)  &   0.04(1) \\
			$L_{3}/L_{1+2+3}$ ($V$)  &   0.06(1) \\
			$L_{3}/L_{1+2+3}$ ($R$)  &   0.09(1) \\
			$L_{3}/L_{1+2+3}$ ($I$)  &   0.14(1) \\
			$e$                    &  \multicolumn{1}{c}{0}
			\\
			$i$ $(^{o})$            &  87.0(2)   \\
			$\Omega_{\rm 1}$       	&  6.919(5)  \\
			$\Omega_{\rm 2}$      	&  4.481(7)  \\
			$\Omega_{\rm cr}$      	&  \multicolumn{1}{c}{4.080(1)} \\
			$ r_{1} $				&  0.1755(6) \\
			$ r_{2} $     		 	&  0.3453(3) \\
			\hline
		\end{tabular}
	\end{center}
\end{table*}

\section{Discussion}

\subsection{Astrophysical parameters and the distance to the system}

The astrophysical parameters of the MQ~Cen components are calculated and presented in Table \ref{table4}. The masses and the radii of the components are derived to within a precision of 1-2\,\% which allowed us to derive their spectral types. Following the table of fundamental stellar parameters given by \cite{straizys}, the mass of the secondary component implies a B4~V-IV star, however, its radius is $\sim$\,35 per cent larger than a normal main-sequence star, indicating an early departure from the main sequence. Nevertheless, the mass and radius of the less massive primary component indicate a typical B5 main-sequence star.

The color excess ($E(B-V)=0.34$ mag) obtained from Str\"{o}mgren colors in \S3 has yielded an estimate of the visual absorption between MQ~Cen and the Sun, as $A_V=0.34\times3.1=1.05$ mag. The absolute visual magnitude of the components was derived by using the bolometric magnitude and correction. The derived distance to MQ~Cen was found to be $d=$\,2.46\,$\pm$\,0.31\,kpc, using the system visual magnitude ($V=10.16$ mag), along with its absolute visual magnitude, visual absorption and the light contribution of the components.

The photometric distance found in this study appears to be slightly larger than the one reported in the \textit{Gaia} Data Release 2 (DR2) Catalogue \citep{gaia} (2.02\,$\pm$\,0.19 kpc), although they are in agreement regarding their uncertainties.

\begin{table*}
	\setlength{\tabcolsep}{3pt}
	\begin{center}
		\caption{Close binary stellar parameters of MQ~Cen. Uncertainty of each parameter is given in brackets.} \label{table4}
		\begin{tabular}{lccc}\hline\hline
			Parameter                 & Symbol  & Primary
			& Secondary     \\
			\hline
%			Spectral type             & Sp      & B4 V     & B5 V      \\
			Mass (M$_\odot$)          & \emph{M}& 4.26 (0.10)  & 5.14 (0.09) \\
			Radius (R$_\odot$)        & \emph{R}& 3.72 (0.05)  & 7.32 (0.03) \\
			Separation (R$_\odot$)    & \emph{a}& \multicolumn{2}{c}{21.2 (0.1)} \\
			Orbital inclination ($^{\circ}$) & \emph{i} & \multicolumn{2}{c}{87.0 (0.2)} \\
			Mass ratio                   & \emph{q} & \multicolumn{2}{c}{1.207 (0.009)} \\
			Eccentricity                 & \emph{e} & \multicolumn{2}{c}{0} \\
			Surface gravity             & $\log g$ & 3.926 (0.015)& 3.420 (0.014) \\
			Color index (mag)          & $B-V$ & \multicolumn{2}{c}{0.17 (0.06)} \\
			Intrinsic colour index (mag)& $(B-V)_{\rm0}$ &\multicolumn{2}{c}{--0.17}\\
			Color excess (mag)         &$E(B-V)$& \multicolumn{2}{c}{0.34}\\
			Visual absorption (mag)     & $A_{\rm v}$ &\multicolumn{2}{c}{1.05 (0.06)} \\
			Int. visual magnitude (mag)  & \emph{$V_{\rm0}$} &  \multicolumn{2}{c}{9.10 (0.09)} \\
			Temperature (K)       & $T_{\rm eff}$  & 16\,600 (520) & 15\,000 (500) \\
			Luminosity (L$_\odot$)      & $\log$ \emph{L}& 2.98 (0.08) & 3.39 (0.66) \\
			Bolometric magnitude (mag)  &$M_{\rm bol}$   & --2.70 (0.20) & --3.72 (0.17) \\
			Abs. visual magnitude (mag)  &$M_{\rm v}$  & --1.22 (0.21) & --2.48 (0.17) \\
			Bolometric correction (mag)   &\emph{BC}& --1.48 (0.07) & --1.24 (0.08) \\
			Velocity amplitudes (km\,s$^{-1}$)&$K_{\rm 1,2}$& 158.9 (0.8) & 131.6 (0.7) \\
			Systemic velocity (km\,s$^{-1}$)  &$V_{\gamma}$ & \multicolumn{2}{c}{--1.2 (4.8)} \\
			Computed sync. vel. (km\,s$^{-1}$)& $V_{\rm synch}$   & 59.7 (0.4) & 101.3 (0.7) \\
			Observed rot. vel. (km\,s$^{-1}$) & $V_{\rm rot}$ & 60 (2) & 100 (5) \\
			Distance (kpc) &\emph{d} & \multicolumn{2}{c}{2.46 (0.31)} \\
			\hline
		\end{tabular}
		\\Note: Solar units were used with following values: M$_\odot$ = 1.9891$\times$10$^{30}$ kg , R$_\odot$ = 695\,990 km, L$_\odot$ = 3.826$\times$10$^{26}$\,W.
	\end{center}
\end{table*}

\subsection{Evolutionary status of MQ~Cen}

\begin{figure*}
	\begin{center}
		\begin{tabular}{c}
			\hspace{1cm}
			\resizebox{100mm}{!}{\includegraphics{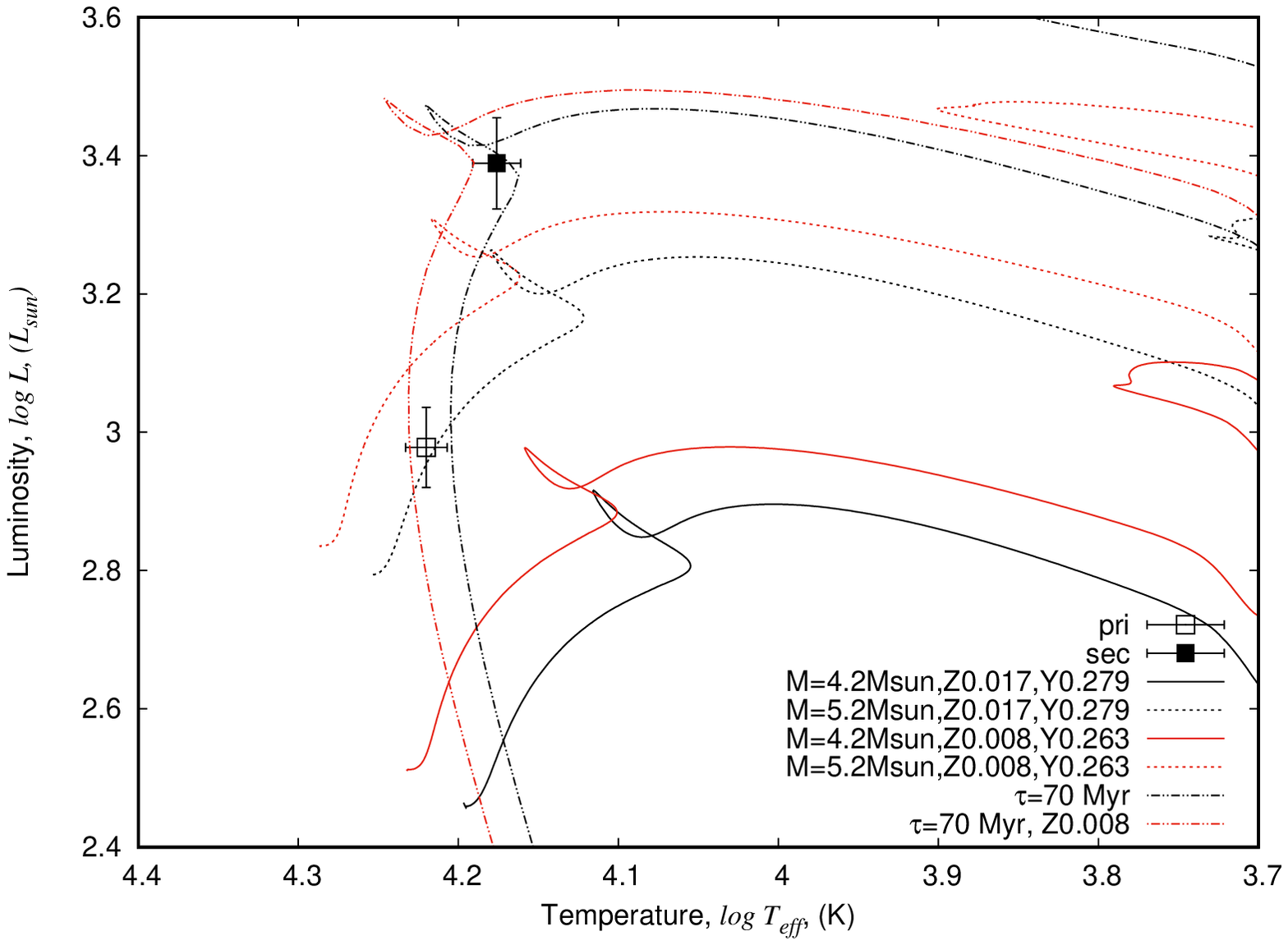}}\\
%			\vspace{0.1cm}
		\end{tabular}
		\caption{Locations of the components of MQ~Cen in the log~$T_{eff}$ -- log~$L$ plane together with PARSEC evolutionary tracks and isochrones for \textit{Z}=0.017, \textit{Y}=0.279 (black) and \textit{Z}=0.008, \textit{Y}=0.263 (red).}\label{fig7}
	\end{center}
\end{figure*}

Each component's evolutionary stage was investigated by locating its position on the HR diagram along with the theoretical evolutionary models (see Fig. \ref{fig7}). The PARSEC models have been adopted from \citet{bressan} for two different compositions (\textit{Z,Y}\,=\,0.008, 0.263 and 0.017, 0.279). The position of the components in the $\log T_{\rm eff} - \log L$ plane indicates that a more massive secondary component has already evolved from the main-sequence, while the primary star currently resides on the main-sequence. Taking into account the uncertainties of the derived parameters, both components lie within the age interval of 70 Myr.

\subsection{Crux OB1 Membership}

The distance to MQ~Cen has been derived to be $d=$\,2.46 $\pm$\,0.31\,kpc in \S4.1, which is in agreement with the distance of the Crux OB1 association (2729 pc) determined by \citet{kaltcheva}. However, MQ~Cen with its age of $\tau_{MQ~Cen}\sim$\,70 Myr seems to be older than the age of the Crux OB1 association ($\tau_{Crux OB1}\sim$\,6.1\,$\pm$\,1.6 Myr \citep{kaltcheva}).

To determine the kinematical properties of MQ Cen, the velocity of the center of mass, the proper motion and the photometric parallax of the system have been used. The center-of-mass velocity ($V_{\gamma}=-1.2\pm4.8$ km s$^{-1}$) and the photometric parallax ($\pi=0.4065\pm0.051$ mas) of the system were evaluated in this study, whereas components of the proper motion ($\mu_{\alpha}\cos\delta$\,=\,--4.874\,$\pm$\,0.078, $\mu_{\delta}$\,=\,0.578\,$\pm$\,0.062 mas yr$^{-1}$) were taken from the {\em Gaia} DR2 Catalogue \citep{gaia}. The space velocity of the system, with respect to the Sun, was calculated using the algorithm described in \citet{johnson} where the space velocity components and their errors were found to be ($U$, $V$, $W$) = (--51.8\,3$\pm$\,6.80, -22.93\,$\pm$\,5.30, --8.10\,$\pm$\,1.26)\,km\,s$^{-1}$. The velocity components were corrected for the peculiar velocity of the Sun, which is known as (8.50\,$\pm$\,0.29, 13.38\,$\pm$\,0.43, 7.00\,$\pm$\,0.26)\,km\,s$^{-1}$ and is given by \citet{coskunoglu}. Therefore, the final Local Standard of Rest (LSR) space velocity components ($U$,$V$, $W$)$_{LSR}$ of MQ~Cen are (--43.33\,$\pm$\,6.81, --9.55\,$\pm$\,5.32, --1.61\,$\pm$\,1.29)\,km\,s$^{-1}$, respectively. The total space velocity of MQ~Cen is $S=44.40\,\pm\,8.74$ km s$^{-1}$. The total space velocity of the system is in agreement with the space velocities of young thin-disc stars \citep{legett}. In addition, the MQ Cen space velocity components obtained in this study, are close to the space velocity components (($U$, $V$, $W$) = (--43.7\,$\pm$\,1.6, -16.6\,$\pm$\,1.9, --6.2\,$\pm$\,0.8) km s$^{-1}$) calculated for the Crux OB1 by \citet{tetzlaff}. Interestingly, if the distance parameter of the system is taken as $\pi=0.4947\pm0.0499$ mas in {\em Gaia} DR2 catalogue \citep{gaia}, then the space velocity components of the MQ~Cen system would become ($U$, $V$, $W$)=(--42.68\,$\pm$\,4.76,--18.65\,$\pm$\,4.79, --6.66\,$\pm$\,0.90)\,km\,s$^{-1}$ which are more compatible with \citet{tetzlaff}. Kinematic analysis shows that MQ~Cen system is strong candidate for membership of the Crux OB1 association.

\section{Concluding Remarks}

Looking at what we have already explained, we conclude our study with the following remarks:

The astrophysical parameters of the massive early-type binary system, MQ~Cen have been determined to a precision of 1-2\,\% which allowed us to construct an evolutionary model for this system. According to the scenario, the more massive secondary component has completed the hydrogen burning phase (main-sequence life). As the secondary component evolves towards the giant branch, it will continue expanding, until reaching its Roche lobe limit in the future. This system is presently located just before the mass transfer stage. Later, it should reach the mass-ratio reversal stage, before becoming a classical Algol-type binary.

According to \citet{zahn}, the components of a binary system become tidally locked in spite of the synchronization of their axial rotation and orbit period. The synchronization time-scale is a function of physical properties of the components and the orbital period of the system. Using the physical properties of the components of MQ~Cen and following the methodology of \citet{zahn}, the synchronization time scale of MQ~Cen was found to be 2~Myr, which indicates that the system has completed synchronization about $\sim$70 Myr ago during the main-sequence lifetime. As expected, the measured rotational velocities of the primary and the secondary components are close to their calculated synchronization velocities (see Table \ref{table4}).

The sky position, the distance and the kinematical parameters of MQ~Cen are in good agreement with those of Crux OB1 association. However, the age of MQ~Cen seems to be larger than the age of Crux~OB1. Being aware of the existence of systems with similar ages would give clues about the existence of a different hierarchical stellar association in the vicinity of Crux~OB1. This would require more comprehensive and systematic observational studies over a relatively long period of time \citep[e.g.][]{pecaut}.

\textbf{Acknowledgements} The spectroscopic observations were performed under ESO grant number {\it 086.D-0236}. The work of ZM, JJ, MZ was supported by GA\,\v{C}R grant 16-01116S. This research has made use of ``Aladin sky atlas'' developed at CDS, Strasbourg Observatory, France. This work has made use of data from the European Space Agency (ESA) mission {\it Gaia} (\url{https://www.cosmos.esa.int/gaia}), processed by the {\it Gaia} Data Processing and Analysis Consortium (DPAC,  \url{https://www.cosmos.esa.int/web/gaia/dpac/consortium}). Funding for the DPAC has been provided by national institutions, in particular the institutions participating in the {\it Gaia} Multilateral Agreement. On the other hand, authors would like to express their warm thanks to Jeremy Tregloan-Reed (CITEVA, University of Antofagasta, Chile) and Ali Akbar Taheri (English Department at Bo\u{g}azi\c{c}i University, Turkey) for their help for correcting the English language and proof reading of the paper.

\end{document}